\documentstyle[amssymb,aps,preprint]{revtex}
\begin{document}
\title{Exact-exchange density-functional theory for quasi-two-dimensional
electron gases}
\author{F. A. Reboredo and C. R. Proetto}
\address{Comisi\'{o}n Nacional de Energ\'{\i}a At\'{o}mica\\
Centro At\'{o}mico Bariloche and Instituto Balseiro\\
8400 Bariloche, Argentina}
\maketitle

\begin{abstract}
A simple exact-exchange density-functional method for a
quasi-two-dimensional electron gas with variable density is presented. An
analytical expression for the exact-exchange potential with only one
occupied subband is provided, without approximations. When more subbands are
occupied the exact-exchange potential is obtained numerically. The theory
shows that, in contradiction with LDA, the exact-exchange potential exhibits
discontinuities and the system suffers a zero-temperature first-order
transition each time a subband is occupied. Results suggesting that the
translational symmetry might be spontaneously broken at zero temperature are
presented. An extension of the theory to finite temperatures allows to
describe a drop in the intersubband spacing in good quantitative agreement
with recent experiments.
\end{abstract}

\newpage

%\twocolumn 
\narrowtext 

Density-functional theory (DFT)\cite{hohen} has been one of the most
successful approaches to the problem of interacting particles. The use of
the local-density-approximation (LDA)\cite{kohn} of DFT and related
approximations has been widely applied to the study of atoms, molecules,
clusters, nuclei, and condensed matter systems.\cite{parr,kohnnobel} At the
same time, steady improvements on the experimental techniques have made
possible the production of a great number of semiconductor artificial
structures that confine the 3D electron gas in one or more dimensions. In
particular{\rm \ GaAs/Al}$_{x}${\rm Ga}$_{1-x}${\rm As} heterostructures
have attracted intense attention, because they are particularly well grown,
their conduction band is simple to describe, and they allow the continuous
change of the electron density. Because typical 3D electron densities in the
quantum well (QW) are very small, the conduction band dispersion can be
approximated, without significant error, by a second order expansion around
the minimum (effective-mass approximation).\cite{ando} As a consequence, the 
{\it effective} electron gas formed in the nanostructure can be treated in
the frame of Kohn-Sham\cite{kohn} DFT. All the theoretical techniques
developed around DFT and its approximations, such as the LDA, time dependent
LDA (TDLDA), spin-density LDA (LSDA) can be tested in these structures which
are much easier to modify and control than natural materials\cite{ando}.
Therefore, semiconductor nanostructures offer the possibility to test over a
broad range of conditions the validity of the theoretical tools that are
used to describe matter in general.

In this paper we present an approach that goes beyond LDA: an exact-exchange%
\cite{xxatoms,xxsemiconductors} density-functional theory for the case of a
quasi-two-dimensional electron gas (2DEG). Because we study a metallic
system with several subbands occupied, our theory is different than previous
exact-exchange (XX)\ procedures aimed to the study of atoms\cite{xxatoms} or
bulk semiconductors\cite{xxsemiconductors}. Moreover, the number of
particles is {\it not} {\it fixed} but it is allowed to fluctuate\cite{parr}%
. When a single subband is occupied we provide an analytical expression for
the XX potential {\it without approximations}. When more subbands are
occupied the XX potential is obtained {\it up to numerical precision}. The
implementation of the method is simple enough to replace the state of the
art method for 2DEG which is currently based in the LDA.\cite{double} We
prove that the XX potential is, in general, a discontinuous function of the
density which can not be accounted by LDA. Therefore, the XX approach yields
qualitatively new physics: at zero temperature a first-order transition
occurs every time a new subband is occupied. In the transition region, the
theory suggest a new phase in which the translational symmetry of the system
might be broken. An extension of the theory to finite temperatures allows us
to describe a drop in the intersubband spacing when the first-subband is
occupied in good quantitative agreement with recent experiments\cite{goni01}.

This theory can be applied to any system with translational symmetry in a
plane. For instance, a QW grown epitaxially as shown schematically in Fig. %
\ref{fg:sys}a. In these systems, it is possible to confine an electron gas
changing the semiconductor in the growth direction $z.$ If the larger gap
semiconductor is doped with donors, it provides electrons to the trap formed
by the smaller gap semiconductor, until the charge-transfer field
equilibrates the chemical potentials in the QW and the doped region (see
Fig. \ref{fg:sys}b ). The charge transferred from the doped reservoir to the
QW can be tuned by an external electric field\cite{goni01}. Assuming
translational symmetry of the 2DEG in the $x-y$ plane (area $A$), and
proposing accordingly a solution of the type $\phi _{i{\bf k}\sigma }\left( 
{\bf r}\right) =\exp (i$ ${\bf k}\cdot {\bf \rho })\xi _{i}^{\sigma }\left(
z\right) /\sqrt{A}$ the ground-state-electron density can be obtained by
solving a set of effective one-dimensional Kohn-Sham equations\cite{kohn} of
the form 
\begin{equation}
\left[ -\frac{1}{2}\frac{\partial ^{2}}{\partial z^{2}}+V_{KS}\left(
z,\sigma \right) \right] \xi _{i}^{\sigma }\left( z\right) =\varepsilon
_{i}^{\sigma }\xi _{i}^{\sigma }\left( z\right) ,  \label{eq:schrod}
\end{equation}%
where $\hbar ^{2}/(m^{\ast }a_{0}^{\ast 2})=2$ Ry$^{\ast }$ and $a_{0}^{\ast
}$ are the units of energy and length, being $m^{\ast }$ the electron
effective mass and $a_{0}^{\ast }$ the effective Bohr radius. $\xi
_{i}^{\sigma }\left( z\right) $ is the wave function corresponding to an
electron with a spin projection $\sigma $ ($\uparrow $ or $\downarrow $),
subband index $i$, and eigenvalue $\varepsilon _{i}^{\sigma }$. In Eq.(\ref%
{eq:schrod}) {\normalsize $V_{KS}(z,\sigma
)=V_{ex}(z)+V_{H}(z)+V_{XX}(z,\sigma )+V_{c}(z,\sigma )$ }is the
spin-dependent Kohn-Sham potential given as sum of the external, Hartree,
local XX, and correlation potentials, respectively. The external potential
is given by the sum of the epitaxial potential plus any external electric
field. The zero-temperature 3D electron density $n(z){}\!$ $%
=\!{}\sum_{\sigma }n(z,\sigma )=\!\sum_{\varepsilon _{i}^{\sigma }<\mu
}\!\!\!\!$ $(\mu -\varepsilon _{i}^{\sigma })\left| \xi _{i}^{\sigma }\left(
z\right) \right| ^{2}/2\pi \!,$ $n(z,\sigma )$ being the fraction of $\sigma 
$ polarized electrons. $\mu $ is the chemical potential (or Fermi level) of
the system, which is determined by the electrostatic and thermodynamic
equilibrium with a reservoir. $V_{H}(z)$ is the solution of the Poisson
equation. We approximate the {\it correlation} potential $V_{c}\left(
z,\sigma \right) $ as a function of the local spin density\cite{perdew}. It
remains to define the XX potential $V_{XX}$.

The exchange energy $E_{X}\left[ n(z,\sigma )\right] $ of a 2DEG can be
obtained through a Slater determinant constructed with the (occupied)
self-consistent solutions of Eq. (1), as follows 
\begin{equation}
E_{X}=-\sum_{i,j,\sigma }k_{F}^{i\sigma }k_{F}^{j\sigma }\!\int \!dz\text{ }%
dz^{\prime }\,\varphi _{i}^{\sigma }(z,z^{\prime })\varphi _{j}^{\sigma
}(z^{\prime },z)F_{ij}^{11}(z,z^{\prime })\!,  \label{eq:energy}
\end{equation}%
where $\varphi _{i}^{\sigma }(z,z^{\prime })=\xi _{i}^{\sigma }(z)^{\ast
}\xi _{i}^{\sigma }(z^{\prime }),F_{ij}^{mn}(z,z^{\prime })=\frac{A}{4\pi }%
\!\int \!\frac{d\rho }{\rho ^{m+n-1}}\frac{J_{m}\!\left( k_{F}^{i\sigma
}\,\rho \right) J_{n}\!\left( k_{F}^{j\sigma }\,\rho \right) }{\sqrt{\rho
^{2}+\left( z-z^{\prime }\right) ^{2}}},$ $k_{F}^{i\sigma }=\sqrt{2}\theta
(\mu -\varepsilon _{i}^{\sigma })\left( \mu -\varepsilon _{i}^{\sigma
}\right) ^{1/2},\theta (x)$ is the step function, and $J_{n}(x)$ stands for
the cylinder Bessel function of order $n$. The XX potential $V_{XX}(z,\sigma
)$ can be obtained from 
\[
V_{XX}(z,\sigma )=\frac{\delta E_{X}}{\delta n\left( z,\sigma \right) }%
=A\sum_{i}\int dz^{\prime }\frac{\delta V_{KS}(z^{\prime },\sigma )}{\delta
n\left( z,\sigma \right) }\times 
\]%
\begin{equation}
\left\{ \int dz^{^{\prime \prime }}\left[ \frac{\delta E_{X}}{\delta \xi
_{i}^{\sigma }\left( z^{^{\prime \prime }}\right) }\frac{\delta \xi
_{i}^{\sigma }\left( z^{^{\prime \prime }}\right) }{\delta V_{KS}\left(
z^{\prime },\sigma \right) }+c.c.\right] +\frac{\delta E_{X}}{\delta
k_{F}^{i\sigma }}\frac{\delta k_{F}^{i\sigma }}{\delta V_{KS}\left(
z^{\prime },\sigma \right) }\right\} .  \label{eq:exact1}
\end{equation}%
The first term in the r.h.s. of Eq.(\ref{eq:exact1}) comes from functional
derivatives with respect to the ``shape'' of the wave function\cite%
{xxsemiconductors}, while the second term is a result of changes in the
occupation of the subbands and has not been considered in previous XX
treatments for fixed particle-number systems. $\delta V_{KS}\left( z^{\prime
},\sigma \right) /\delta n\left( z{\bf ,}\sigma \right) \equiv \chi _{\sigma
}^{-1}(z,z^{\prime })$ is the inverse of the operator $\chi _{\sigma
}(z,z^{\prime })\equiv \delta n\left( z{\bf ,}\sigma \right) /\delta
V_{KS}\left( z^{\prime }\right) $ given by 
\begin{equation}
\chi _{\sigma }(z,z^{\prime })=\sum_{i}\left\{ \frac{\left( k_{F}^{i\sigma
}\right) ^{2}}{4\pi A}\left[ \varphi _{i}^{\sigma }(z,z^{\prime })G_{i\sigma
}(z,z^{\prime })+c.c.\right] -\frac{f(\varepsilon _{i}^{\sigma })}{2\pi A}%
\left| \varphi _{i}^{\sigma }(z,z^{\prime })\right| ^{2}\right\} ,
\label{eq:xhi}
\end{equation}%
being $G_{i\sigma }(z,z^{\prime })=\sum_{j(\neq i)}\varphi _{j}^{\sigma
}(z^{\prime },z)/(\varepsilon _{i}^{\sigma }-\varepsilon _{j}^{\sigma }),$
and $f(\varepsilon _{i}^{\sigma })=1/\left[ 1+\exp ((\varepsilon
_{i}^{\sigma }-\mu )/k_{B}T))\right] $ the Fermi occupation factor\cite%
{sommerfeld}. The first term in Eq.(\ref{eq:xhi}) comes from first-order
perturbation theory\cite{gorling}, whereas the second term results from
first-order perturbation theory and the thermodynamic equilibrium between
the 2DEG and the reservoir that fixes a common chemical potential $\mu $
allowing the change of the number of particles. Indeed, without this second
term, the operator $\chi _{\sigma }(z,z^{\prime })$ in general cannot be
inverted because it is singular\cite{xxatoms,xxsemiconductors}. Eq.(\ref%
{eq:exact1}) could be brought to an alternative expression, which allows the
discussion of our approach in the context of the Optimized Potential Method
(OPM).\cite{OEP} For this, we multiply Eq.(\ref{eq:exact1}) by $\chi
_{\sigma }(z,z^{\prime })$, and integrate over all $z.$ Proceeding this way,
it could be rewritten as 
\begin{eqnarray}
&&\sum_{i}(k_{F}^{i\sigma })^{2}\int dz^{\prime }\left\{ \left[
V_{XX}(z^{\prime },\sigma )-v_{X,i}(z^{\prime },\sigma )\right] G_{i\sigma
}(z^{\prime },z)\varphi _{i}^{\sigma }(z^{\prime },z)\right.   \label{eq:OEP}
\\
&&\left. +c.c.\right\} -2\sum_{i}^{occ}\left| \xi _{i}^{\sigma }(z)\right|
^{2}\left[ \overline{V}_{XX,i}(\sigma )-\frac{2\pi }{Ak_{F}^{i\sigma }}\frac{%
\delta E_{X}}{\delta k_{F}^{i\sigma }}\right] =0.  \nonumber
\end{eqnarray}%
Here, $v_{X,i}(z,\sigma )=\left[ 4\pi /A(k_{F}^{i\sigma })^{2}\xi
_{i}^{\sigma }(z)^{\ast }\right] \delta E_{X}/\delta \xi _{i}^{\sigma }(z),$
and $\overline{V}_{XX,i}(\sigma )$ is the diagonal matrix element of $%
V_{XX}(z,\sigma )$ with $\xi _{i}^{\sigma }(z).$ This integral equation for $%
V_{XX}(z,\sigma )$ is the generalization of the OPM to our open
configuration, where the system is free to  exchange particles with the
surroundings; for a closed system (fixed number of particles), the last term
on the r.h.s. of Eq.(\ref{eq:OEP}) is zero, and the integral equation
reduces to the one of the standard OPM for atoms and molecules.\cite{OEP}
Some consequences of Eq.(\ref{eq:OEP}) are worth of address: {\it a)} the
solution for $V_{XX}(z,\sigma )$ is {\it univocally} determined, including
the (possible) presence of a constant shift $C$; {\it b)} its solution in
the one-subband case $i=0$ is immediate: Replacing $V_{XX}^{0}(z,\sigma
)=v_{X,0}(z,\sigma )+C$ in Eq.(\ref{eq:OEP}) and solving for $C,$ we found 
\begin{equation}
V_{XX}^{0}(z,\sigma )=-\frac{8\pi }{A}\left\{ \int dz_{1}\varphi
_{0}^{\sigma }(z_{1},z_{1})F_{00}^{11}(z_{1},z)+k_{F}^{0\sigma }\int
dz_{1}dz_{2}\left| \varphi _{0}^{\sigma }(z_{1},z_{2})\right| ^{2}\left[ 
\frac{F_{00}^{01}(z_{1},z_{2})}{2}-\frac{F_{00}^{11}(z_{1},z_{2})}{%
k_{F}^{0\sigma }}\right] \right\} .  \label{eq:onepot}
\end{equation}%
Using Eq.(\ref{eq:onepot}) it can be shown that $\int V_{XX}^{0}(z,\sigma )%
\left[ \xi _{0}^{\sigma }\left( z\right) \right] ^{2}=dE_{X}/dN_{\sigma
}=\mu _{X}^{\sigma }$, with $E_{X}$ given by Eq.(\ref{eq:energy}), $%
N_{\sigma }=\int d{\bf r}$ $n(z,\sigma )$ the total number of electrons with
spin $\sigma ,$ and $\mu _{X}^{\sigma }$ the exchange contribution to the
total chemical potential $\mu .$ Accordingly, the Janak theorem\cite{janak}
is explicitly satisfied in our XX-DFT formalism. Besides, when Eq.(\ref%
{eq:onepot}) is particularized to the strict 2D limit by imposing the
condition $\left[ \xi _{0}^{\sigma }\left( z\right) \right] ^{2}\rightarrow
\delta (z)$ [$\delta (z)$ being the Dirac delta function], we obtain $%
V_{XX}^{0}(0,\sigma )=-2k_{F}^{0\sigma }/\pi ,$ which is exactly the well
known value of the exchange contribution to the chemical potential for an
homogeneous 2D electron gas. In other words, Eq.(\ref{eq:onepot}) contains
the exact 2D limit of the exchange potential.\cite{kim} Finally c){\it ,}
for two (or more) occupied subbands, the analytical exact solution of our
OPM integral equation is unknown. Some progress can be achieved following
the Sharp-Horton (SH) or KLI approximation\cite{OEP} of the orbital Green
function $G_{i\sigma }(z,z^{\prime })$. It amounts to replace the
denominators $(\varepsilon _{i}^{\sigma }-\varepsilon _{j}^{\sigma })$ by an
orbital independent average $\Delta \widetilde{\varepsilon }_{\sigma };$
substituting $G_{i\sigma }(z,z^{\prime })\simeq \left[ \delta (z-z^{\prime
})-\sum_{i}\left| \varphi _{i}^{\sigma }(z,z^{\prime })\right| ^{2}\right]
/\Delta \widetilde{\varepsilon }_{\sigma }$ in Eq.(\ref{eq:OEP}) one obtains
an explicit solution for $V_{XX}(z,\sigma )$, which will be given elsewhere.%
\cite{else} For the one-subband case, the KLI approximation leads to the
exact result, given by Eq.(\ref{eq:onepot}). But for the multi-subband case $%
(i\geqslant 1)$, and because of the second term in Eq.(\ref{eq:OEP}), the
explicit solution for $V_{XX}(z,\sigma )$ becomes $\Delta \widetilde{%
\varepsilon }_{\sigma }$ dependent, in contrast with the situation for atoms
and molecules.\cite{OEP}

As proceeding along this line of work would had forced us to introduce a new
and unknown scale of energy $(\Delta \widetilde{\varepsilon }_{\sigma })$,
we studied the multi-subband case by directly solving Eq.(4). \ The
fundamental ingredient for this direct approach is an analytical limit for $%
\chi _{\sigma }^{-1}\left( z,z^{\prime }\right) $. For $T=0$ and in the case
where {\it only} the ground state subband is occupied $\chi _{0\sigma
}^{-1}\left( z,z^{\prime }\right) =\sum_{i}d_{i}^{\sigma }\psi _{i}^{\sigma
}\left( z\right) \psi _{i}^{\sigma }\left( z^{\prime }\right) ,$where $%
d_{0}^{\sigma }=-2$ , $d_{i>0}^{\sigma }=-(\varepsilon _{i}^{\sigma
}-\varepsilon _{0}^{\sigma })/(\mu -\varepsilon _{0}^{\sigma }),$ and $\psi
_{i}^{\sigma }\left( z\right) =\xi _{i}^{\sigma }\left( z\right) /\xi
_{0}^{\sigma }\left( z\right) .$ Replacing this and Eq.(\ref{eq:energy})
into (\ref{eq:exact1}) we get again, by a different method, Eq.(\ref%
{eq:onepot}). For many occupied subbands, one can obtain $V_{XX}(z,\sigma )$
using Eq.(\ref{eq:exact1}) because $\chi _{\sigma }^{-1}(z,z^{\prime })$ can
be obtained recursively in terms of $\chi _{0\sigma }^{-1}\left( z,z^{\prime
}\right) $ for any number of bands and temperature, by using the
Sherman-Morrison method.\cite{SM} Thus for more than one subband we have
evaluated $V_{XX}(z,\sigma )$ up to numerical precision. In order to test
the numerical method we verified that, in double quantum well systems with 
{\it two} occupied subbands, the {\it single} subband analytical limit for
the XX potential in one well (Eq.(\ref{eq:onepot})), is nicely reproduced
numerically as we increase the barrier between the wells.

Fig.\ref{fg:sys}b gives the full self-consistent potential $V_{KS}(z)$ and
the squared wave functions $\left[ \xi _{i}^{\sigma }\left( z\right) \right]
^{2}$ corresponding to the first four states. The results were obtained for
a QW of {\rm GaAs} in {\rm Al}$_{x}${\rm Ga}$_{1-x}${\rm As} with a band
offset of $220$\ meV and a width of $245$ \AA . The structure is doped on
one side at $135$ \AA\ from the left QW edge with a 3D dopant density of $%
2\times 10^{18}/$cm$^{-3}$. Without external electric field and at zero
temperature the ground state is paramagnetic, the intersubband spacings are $%
E_{01}=\varepsilon _{1}^{\sigma }-\varepsilon _{0}^{\sigma }$ $=26.78$ meV, $%
E_{02}=60.58$ meV, $E_{03}=108.36$ meV, while $\mu -\varepsilon _{0}^{\sigma
}$ $=24.14$ meV is just below $\varepsilon _{1}^{\sigma }-\varepsilon
_{0}^{\sigma }$ and only one subband is occupied.

In Fig.\ref{fg:potcomp} we compare the potential $V_{XX}^{0}(z,\sigma )$
corresponding to Fig.\ref{fg:sys}b  [obtained using Eq.(\ref{eq:onepot})]
with the exchange potential in LDA $V_{x}^{LDA}(z,\sigma )=-\left[
6n(z,\sigma )/\pi \right] ^{1/3}$ for the {\it same} XX density. Note that
although $V_{x}^{LDA}$ and $V_{XX}^{0}$ have a similar amplitude in the QW
region, their respective asymptotic behaviors are completely different:
while $V_{x}^{LDA}$ goes exponentially to zero for $\left| z\right|
\rightarrow \infty ,$ $V_{XX}^{0}$ tends asymptotically (as $-1/\left|
z\right| $) towards the constant (positive) contribution of Eq.(\ref%
{eq:onepot}). This results in larger intersubband spacings using XX theory
for the same total electron density. Note that the minimum of $%
V_{XX}^{0}(z,\sigma )$ is not at the maximum of $n\left( z,\sigma \right) $
as in $V_{x}^{LDA}(z,\sigma )$ and also note that $V_{XX}^{0}(z,\sigma )$ is
large where $n\left( z,\sigma \right) $ is negligible. Both features are
consequences of the non-local dependence of $V_{XX}^{0}(z,\sigma )$ on $%
n\left( z,\sigma \right) ,$ displayed explicitly in Eq.(6)$.$

Significant qualitative differences between LDA and XX theory appear each
time a new subband is occupied. For instance in Eq.(\ref{eq:exact1}), at
zero temperature, a finite term proportional to $\delta E_{X}/\delta
k_{F}^{1\sigma }$ appears discontinuously\cite{goni01} when $k_{F}^{1\sigma
}\rightarrow 0^{+}$. Moreover, in Eqs.(\ref{eq:xhi}) and (\ref{eq:OEP}),
when $\mu \rightarrow \varepsilon _{1}^{\sigma }+0^{+},$ a finite
contribution $(-\left| \varphi _{1}^{\sigma }(z,z^{\prime })\right|
^{2}/2\pi A)$ appears in the second term. Therefore, the inverse $\chi
_{\sigma }^{-1}\left( z,z^{\prime }\right) $ also \ changes discontinuously
when $k_{F}^{1\sigma }\rightarrow 0^{+}$. Although the discontinuities in $%
\delta E_{X}/\delta k_{F}^{1\sigma }$ and $\chi _{\sigma }^{-1}$ have
opposite effects, they do not cancel each other and $\chi _{\sigma }^{-1}$
dominates\cite{else}. Let us consider what happens when $\mu $ crosses a
subband energy $\varepsilon _{n}^{\sigma }$ but one neglects the
self-consistent adjustments of the charge. In that case, because of the
discontinuities introduced by $\delta E_{X}/\delta k_{F}^{1\sigma }$ and in $%
\chi _{\sigma }^{-1}$, the XX potential must change discontinuously: 
\begin{equation}
V_{XX}(z,\sigma ,\mu \rightarrow \varepsilon _{n}^{\sigma
}+0^{+})=V_{XX}(z,\sigma ,\mu \rightarrow \varepsilon _{n}^{\sigma
}-0^{+})+\Delta V_{XX}^{n-1,n}(z,\sigma ).
\end{equation}%
Therefore, the wave functions, the total electron density, and the total
energy cannot be continuous functions of $\mu $ and the discontinuity $%
\Delta V_{XX}^{n-1,n}$ signals a first-order transition of the 2DEG. An
interesting question is what is the effect of $\Delta V_{XX}^{0,1}(z,\sigma )
$ on the intersubband spacing $E_{01}$. Case I: if one only considers the
discontinuities in $\delta E_{X}/\delta k_{F}^{1\sigma }$, $E_{01}(\mu
\rightarrow \varepsilon _{1}^{\sigma }+0^{+})<$ $E_{01}(\mu \rightarrow
\varepsilon _{1}^{\sigma }-0^{+})$  and $\mu $ lies above the first-excited
subband and a {\it finite} amount of charge is abruptly transferred from the
ground to the first-excited subband. Case II: if one adds the
discontinuities in $\chi _{\sigma }^{-1},$    $E_{01}(\mu \rightarrow
\varepsilon _{1}^{\sigma }+0^{+})>$ $E_{01}(\mu \rightarrow \varepsilon
_{1}^{\sigma }-0^{+})$ and $\mu $ remains below the bottom of the
first-excited subband and a self-consistent solution of the system is not
possible (under our assumption of translational symmetry). In words, if the
system attempts the occupation of the first-excited subband, this increases
the intersubband spacing, which in turn empties the excited subband in the
next iteration, which in turn produces a lower intersubband spacing, and so
on. Finally, if $E_{01}(\mu \rightarrow \varepsilon _{1}^{\sigma
}+0^{+})=E_{01}(\mu \rightarrow \varepsilon _{1}^{\sigma }-0^{+})$ (Case
III) a smooth occupation of the first-excited subband would be possible.
Provided that self-consistency is frozen, the only possibility in the LDA is
Case III, because the LDA exchange potential is a continuous function of the
density. Self-consistent effects, in some conditions, can generate
first-order transitions within LDA\cite{double}. But in XX theory, in
contrast, a second order transition will be an accident and, in general,
first-order transitions are expected when $\mu $ crosses a subband energy at
zero temperature.\cite{takada}

In Fig.\ref{fg:spacing} we plot the intersubband spacing $E_{01}$ as a
function of the total 2D electron density $N/A=(N_{\uparrow }+N_{\downarrow
})/A$ with different methods. The dotted line corresponds to the usual LDA,
the continuous line to XX theory at zero temperature, and the dashed line
was obtained with an approximated\cite{sommerfeld} finite temperature XX
theory ($T=10$ K in Eq.(\ref{eq:xhi})). The straight line is the evolution
of $\mu -\varepsilon _{0}^{\sigma }=(\pi \hbar ^{2}/m^{*})N/A$ in an
hypothetical system with a single band occupied. All methods give similar
qualitative results when $E_{01}$ is far from $\mu -\varepsilon _{0}^{\sigma
}$, though LDA calculations give smaller intersubband spacings than XX
methods (as expected from Fig.\ref{fg:potcomp}). However, significant
qualitative differences appear when $E_{01}$ $\approx $ $\mu -\varepsilon
_{0}^{\sigma }.$ The LDA calculation gives a continuous curve but there is a
discontinuity in the derivative of $E_{01}$ when the first-excited subband
is occupied (which implies a second order transition). The zero temperature
calculations with XX theory give a remarkable result: there is a window of
densities where a self-consistent solution is not achieved because it
corresponds to the Case II discussed above. Our interpretation of this
window, where it is not possible to find a self-consistent solution under
the assumption of translational invariance, is that this symmetry must be
broken in the ground state. Thus in this density region a new broken
symmetry phase might exist at low temperatures.

Consideration of finite temperatures in Eq.(\ref{eq:xhi}) introduces
smoothly the discontinuity in $\chi _{\sigma }^{-1}\left( z,z^{\prime
}\right) $ when $\left| \mu -\varepsilon _{1}^{\sigma }\right| \approx
k_{B}T.$ This allows us to achieve a self-consistent solution for all
densities (although the numerical convergence becomes unstable in some
cases). As the density increases, the occupation of the first-excited
subband cannot be avoided and $\delta E_{X}/\delta k_{F}^{1\sigma }$
generates a sudden drop in $E_{01}$. This new solution is also stable at
lower densities. Accordingly, there is a range of densities where it is
possible to find {\it two} solutions. A first-order transition occurs when
the free energies of the two solutions cross; detailed temperature-dependent
calculations will be reported elsewhere. From Fig.\ref{fg:spacing} we can
estimate the drop on $E_{01}$ considering the energies in the density range
where the two solutions exist: $\Delta E_{01}=2.4\pm 0.4$ meV, which is in
very good agrement with the measurement of 3.5 meV reported by Go\~{n}i {\it %
et al.}\cite{goni01}.

In summary we have extended the KS-DFT for 2DEG beyond the state of the art
method based on the LDA. Our theory allows to obtain an exact exchange
potential without approximations up to numerical precision, while
correlations are considered in the LDA level. The theory satisfies known
limits and theorems. The theory predicts first-order transitions every time
a new subband is occupied and suggest that for some systems at zero
temperature the translational symmetry might be spontaneously broken. Finite
temperature effects are included approximately. We calculated the
self-consistent solutions for a realistic system and obtained phase
transitions which are in good quantitative agreement with recent experiments.

The authors would like to thank A. Go\~{n}i for the experimental information
and discussions that inspired this work, and to V. H. Ponce for useful
discussions and a critical reading of the manuscript. We are indebted to
CONICET and Fundaci\'{o}n Antorchas of Argentina for financial support.

\begin{figure}[tbp]
\caption{ (a) Schematic representation of an asymmetric doped quantum well
nanostructure. (b) Self-consistent subband structure and self-consistent
potential for the zero-bias situation. }
\label{fg:sys}
\end{figure}

\begin{figure}[tbp]
\caption{ Comparison of the exact exchange potential with the LDA potential
when only one subband is occupied. The full upper line is the 3D density $%
n(z)$ obtained with XX theory (corresponding to a 2D density of $0.68\times
10^{12}/$cm$^{2}$). XX and LDA exchange potentials are compared for this
same density. Energies are measured from the chemical potential $\protect\mu %
.$}
\label{fg:potcomp}
\end{figure}

\begin{figure}[tbp]
\caption{ Intersubband spacing $E_{01}$ as a function of the 2D density $%
N/A. $ Full lines, zero temperature exact exchange; dashed lines, finite
temperature exact exchange; dotted line, LDA. }
\label{fg:spacing}
\end{figure}


\begin{references}
\bibitem{sommerfeld} Note that we have introduced the temperature only in $%
\chi _{\sigma }\left( z,z^{\prime }\right) $ $($and consequently in $\chi
_{\sigma }^{-1}\left( z,z^{\prime }\right) )$ but not in $E_{X}$. This
approximation (temperature independent term in a Sommerfeld expansion) will
be valid when $\left| \mu -\varepsilon _{i}^{\sigma }\right| >k_{B}T$ for
all $\varepsilon _{i}^{\sigma }$.
\end{references}
\end{document}